\documentclass[aps,pre,twocolumn,groupedaddress]{revtex4-1}
\usepackage{epsfig}
\usepackage{amsmath,amssymb}
\usepackage{color}

\newcommand{\rev}[1]{{\color{black}#1}}
\newcommand{\revtwo}[1]{{\color{black}#1}}

\begin{document}

%\preprint{}

\title{Collective intelligence: aggregation of information from neighbors in a guessing game}


\author{Toni P\'erez}
\email[]{toni@ifisc.uib-csic.es}
%\thanks{}
\author{Jordi Zamora}
\author{V\'ictor M. Egu\'iluz}
\affiliation{Instituto de Física Interdisciplinar y Sistemas Complejos IFISC 
(CSIC-UIB). E07122 Palma de Mallorca, Spain.}


%\date{\today}

\begin{abstract}
Complex systems show the capacity to aggregate information and to display 
coordinated activity. In the case of social systems the interaction of \rev{different} 
individuals leads to the emergence of norms, trends in political positions, 
opinions, cultural traits, and even scientific progress. Examples of collective 
behavior can be observed in activities like the Wikipedia and \rev{Linux}, 
where individuals aggregate their knowledge for the benefit of the 
community, and citizen science, where the potential of collectives to solve 
complex problems is exploited. Here, we \revtwo{conducted} an online experiment to 
investigate the performance of a collective \rev{when solving} a guessing problem \rev{in which} 
each actor is endowed with partial information \rev{and placed as the nodes of an} interaction 
network. We measure the performance of the collective in terms of \rev{the temporal} 
evolution of the \rev{accuracy, finding no} statistical difference in the performance for two classes of networks, regular 
lattices and random networks. We also \rev{determine that a Bayesian description captures the behavior \revtwo{pattern} the individuals follow} \revtwo{in} aggregat\revtwo{ing} 
information from neighbors \rev{\revtwo{to} make decisions. \revtwo{In} comparison with other simple decision models, the strategy followed by the players} reveals a 
suboptimal performance of the collective. Our contribution \rev{provides the basis for} the 
micro-macro connection between individual based descriptions and collective \rev{phenomena}.
\end{abstract}

%\pacs{}
%\keywords{}

\maketitle

\section*{Introduction}
Decision making nowadays is a topic \revtwo{of} growing interest \revtwo{from} \rev{the} scientific 
community and the industry. \revtwo{This is because,} it can provide statistical predictions of 
animal and human behavior \cite{Sumpter2009,King2007,Arganda2012,Nishi2013,Perc2013,Lambiotte2008}. 
\rev{The effect\revtwo{s} of social influence and collaboration on the collective outcome have been addressed from different perspectives, both theoretical and experimental  
\cite{Suri2011,Centola2011,Moussaid2013,Mavrodiev2013,Miller2013}.} 
In complex problem solving, crowd-sourced collaboration has proved \rev{to be beneficial by shortening} 
substantially the time to find a solution. Examples of these collaborative 
initiatives \revtwo{include} problem solving in mathematics \cite{Polymath2009}, software design \cite{TopCoder2010}, 
and data analysis \cite{Kaggle2013}. \\

\rev{In the study of cooperation, network reciprocity is an important mechanism  \cite{Nowak1992,KleinberBook}. 
 Theoretical works have shown that the role of network connectivity on cooperation depends on the evaluation of the individuals payoffs 
 that can have a positive effect \cite{Pacheco2006,Nowak2006}, however, cooperation can be diminished when the real connection cost is take\revtwo{n} into account \cite{Masuda2007}. 
 Experimental work in the Prisoner’s dilemma game shows that heterogeneous networks do not 
enhance cooperation \cite{Gracia2012}. However, a dynamic interaction network can promote cooperation \cite{Rand2011,Eguiluz2005}. 
Other works have also addressed how individuals learn and how the interaction network can influence the dissemination of useful information, 
individual choices, and the social outcomes \cite{Bala1998,Goyal2011}.}
The performance of groups when compared with the performance of individual experts is reflected in The Wisdom of Crowds effect \cite{Galton1907,Surowiecki2004} 
\revtwo{which suggests that the aggregation of independent decisions often outperforms individual experts.}
%reflecting that the estimate of a group of independent individuals can be more accurate than appraises of individual experts. 
However, social influence can have a diminishing effect over this phenomen\revtwo{on} \cite{Lorenz2011}. \\

\revtwo{To investigate how network structure affects social information use, decision making and decision accuracy, we performed an experiment with neutral items
where individuals have to assemble information from peers according to an interaction network.}
% To address the influence of the network in the decision making, we perform an experiment with neutral items
% where individuals have to assemble information from peers according to an interaction network.
Players have to make decisions based on incomplete and uncertain information, which allow us to address the process of 
aggregation of information and assess the decision making \revtwo{capacities} of the players.

\section*{Materials and Methods}

\subsection*{Ethics Statement}

This study was approved by the Ethics Committee of the University of the 
Balearic Islands. Online informed consent was obtained from each participant 
prior to participation in the experiment.

\subsection*{Experimental Design}
\begin{figure}[ht]
\includegraphics[width=\linewidth]{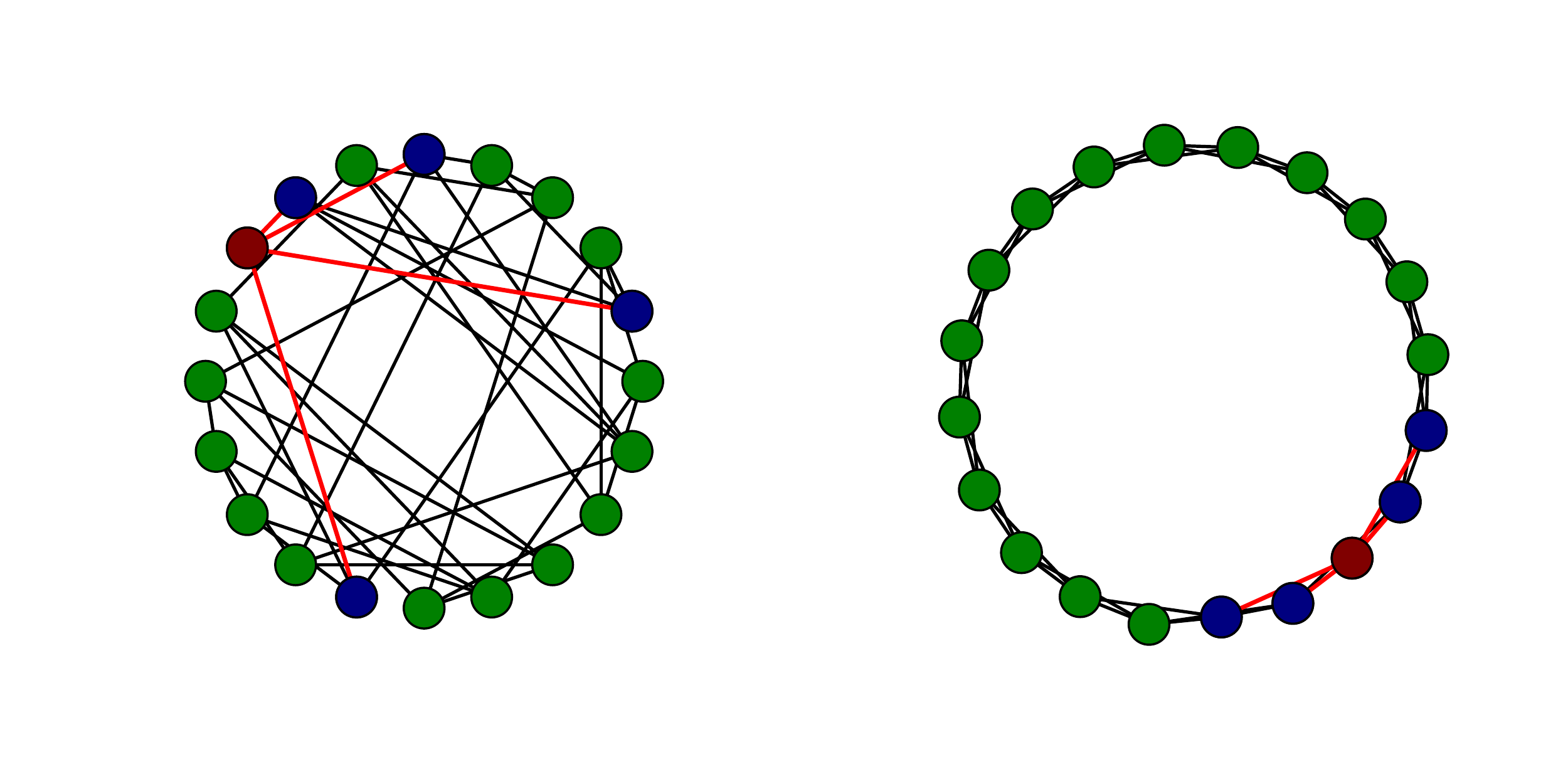}
% %\vspace{3cm}
\caption{{\bf Illustration of the network configurations.} 
 Focal player (red node) is connected to k=4 neighbors (blue nodes) 
 having access to their proposals, and, at the same time, she shares her proposals with them. 
 The remaining nodes and network connections are depicted by the green nodes and the gray links, respectively. 
 Random network is represented on the left and regular network on the right.}
 \label{fig1}
\end{figure}

We developed a social experiment consisting \revtwo{of} an online game in which players 
have to guess a sequence of colors using information of an incomplete sequence 
provided initially to them and from the proposals of their neighbors. 
The experiment \revtwo{was} structured in sessions each consisting \revtwo{of} a set of $N$ individuals 
assigned randomly to the nodes of a network as sketched in Fig. \ref{fig1}. The target 
color code \revtwo{was} composed of a sequence of $l_i$ positions ($i=1,...,10$) colored 
with color $c(i)$ from the available set (red, blue, and yellow). We defined 
$x_j(l_i,t)$ as the color chosen by player $j$ for the position $l_i$ at time $t$.

We conducted a series of online sessions on different dates. The first one \revtwo{took place} 
on May 7th, 2014 at 12:30 GMT+1 (Experiment 1) with 20 participants and a 
second experiment \revtwo{was conducted} 3 weeks later on May 28th, 2014 at 11:30 GMT+1 
(Experiment 2) with 17 participants. \rev{Both experiments were announced through a mailing list, 
and the \revtwo{players\textquoteright~} participation was voluntary and anonymous.} Prior to the start of the experiment\revtwo{,} the participants 
were required to agree on the Terms and Conditions and \revtwo{they were} invited to answer a survey providing 
basic demographic data. The participants, $74\%$ males and $26\%$ females, reported 
an average age of $33.6\pm 7.3$ years. \\
Each experiment consist\revtwo{ed of six} consecutive games being 
the first game intended to familiarize the participant with the interface and 
the \revtwo{purpose of the} last game to evaluate the adaptation process \revtwo{of the players}. The first \revtwo{and} the 
last game\revtwo{s were} played against automata. In the remaining games, the participants 
interact\revtwo{ed} with other participants. Players \revtwo{were} not informed whether they played 
with automata or \revtwo{with} human players. 
An itemized list of instructions \revtwo{was} provided at the beginning of each experiment. During the game, the 
participants also ha\revtwo{d} access to a summary of the main instructions available on the screen 
(see Supporting Information for a description of the instructions provided). No 
economical incentive was offered to the participants. 

\begin{figure}[ht]
\includegraphics[width=\linewidth]{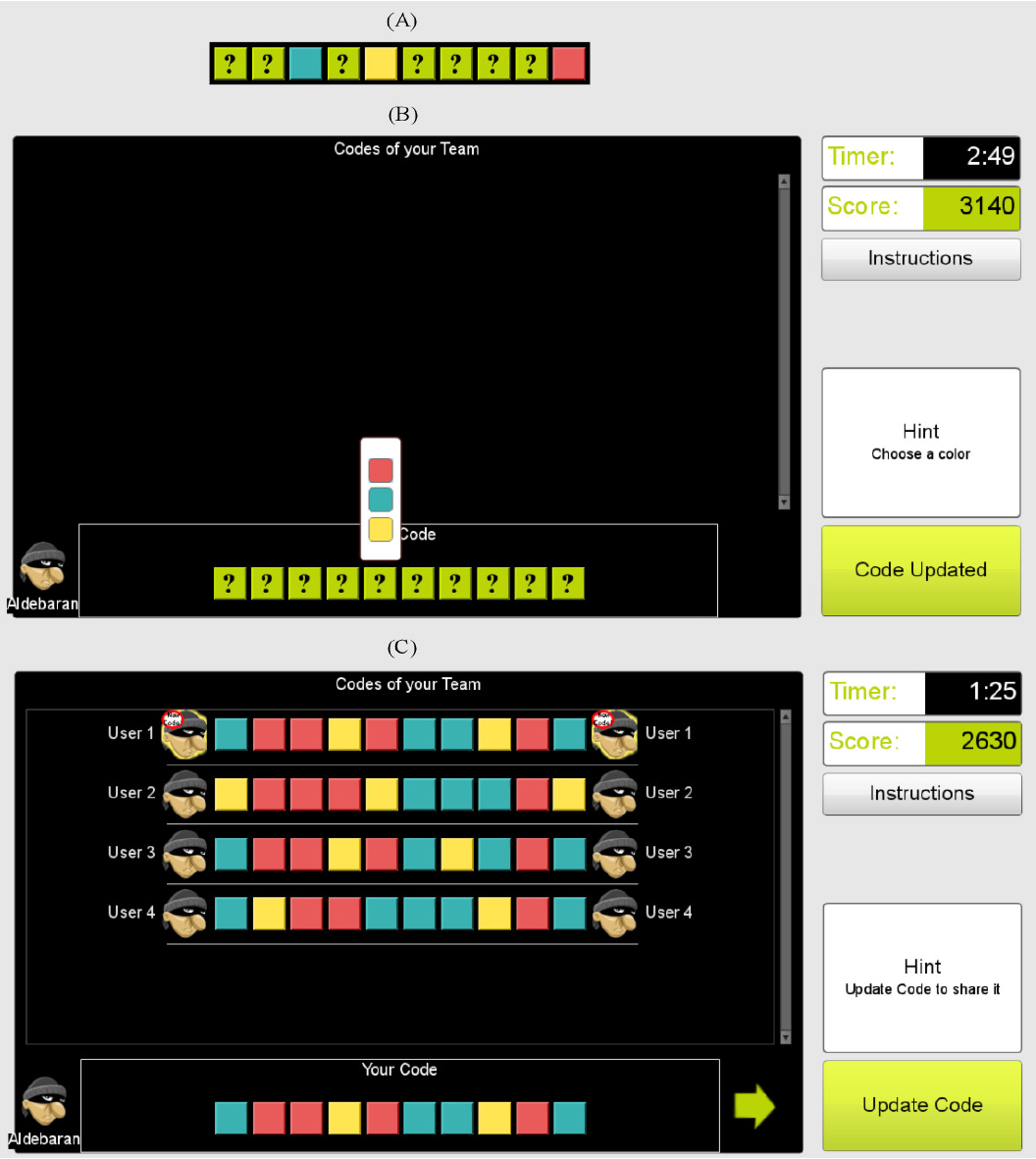}
% \vspace{2cm}
\caption{\rev{{\bf Interactive interface of the game.} 
(A) Example of the partial target code shown initially to the player displaying positions 3, 5 and 10 with the correct color, while the other positions are empty. 
Green boxes with question marks are the remaining positions of the code to be guessed by the player (green color is not used in the color code sequence). 
(B) Interface for the first guess. When the mouse cursor is placed over a question mark green box, a list of available colors present in the code is shown. 
(C) Player interface during the game. At the top of the screen the proposals from neighbor\revtwo{ing} players (Team) are shown. When a member of the Team update\revtwo{s} her code, 
a flashing 'new code' bubble is shown on the icon of the updater.}}
\label{fig2_panel}
\end{figure}

In each game, players ha\revtwo{d} to guess a \rev{color} code of ten 
positions. \rev{Each position \revtwo{could} be colored with one of the following colors: red, blue, or yellow.} 
Participants ha\revtwo{d} \rev{225 seconds} to guess the code \rev{after initial access to a} partial sequence of 
\rev{the code}. An example \revtwo{as well as} the proposed \rev{codes} of the 
neighboring players is shown in Fig. \ref{fig2_panel} A.

\rev{At the beginning of the game, only three out of the ten positions of the common target color code \revtwo{were} shown to the players.  
The positions shown \revtwo{were} generated at random covering the full code and they \revtwo{were} randomly assigned to the players.}
After \revtwo{seeing} the initial partial \rev{code, the players} face\revtwo{d} \rev{an empty} code \rev{(}green boxes 
with question marks\rev{)} that they had to fill with their proposals. When the cursor \revtwo{was} 
placed over one of the positions, a list of available colors \revtwo{was} displayed on 
top of it (see Fig. \ref{fig2_panel} B). To avoid any bias towards any particular color, the 
order of the \rev{list} of available colors presented to each player \revtwo{was} randomly 
generated. When \revtwo{a} player \revtwo{choses} one of the colors \revtwo{in} the list, the position 
is colored accordingly. \rev{Once the player fillS the entire color code}, the 
update button is activated to submit and share the proposed color code. 
\rev{The player \revtwo{could then} see the proposals of the neighbors only after the completion of the first guess.}

Each player \revtwo{was} initially randomly assigned to \revtwo{a} node \revtwo{in the} network and \revtwo{was} 
limited to share \revtwo{her} proposals only with \revtwo{the} neighbors. Consequently,
players only \revtwo{saw} the proposals of their neighbors. Fig. \ref{fig2_panel} C shows 
the game interface after the first proposal \revtwo{of} the \revtwo{focal} player now including the 
proposals of the neighbors. 
\rev{The visualization interface limits the size of the neighborhood, and, to simplify the analysis, we consider\revtwo{ed} networks where each node ha\revtwo{d} the same number of neighbors. 
In particular we use\revtwo{d} a regular lattice and \revtwo{a} degree regular random network as they show differentiated topological properties, for instance\revtwo{,} with respect to the average path length.} 
In order to avoid the effect of learning, if any, the topological 
arrangement \revtwo{of the} networks in Experiment 1 was random, regular, regular, and random, \revtwo{while} in 
Experiment 2 \revtwo{it} was permuted to regular, random, random, and regular. In all 
network configurations, the node degree was set to $k=4$. \\
At the end of each game, 
the players \revtwo{were} rewarded with 10 points for finishing 
the game, and 100 points per correct position. A bonus of 1000 points \revtwo{was} 
granted to the players guessing the correct code. \rev{Before starting a new} game, a ranking of 
the players \revtwo{was} shown. \rev{This \revtwo{was} shown as an incentive to the players \revtwo{although but} it \revtwo{was} not used in the analysis.} 
%A movie with the dynamics of one of the games can be found in the Supporting Information. 

\section*{Results}

\begin{figure}[ht]
\includegraphics[width=\linewidth]{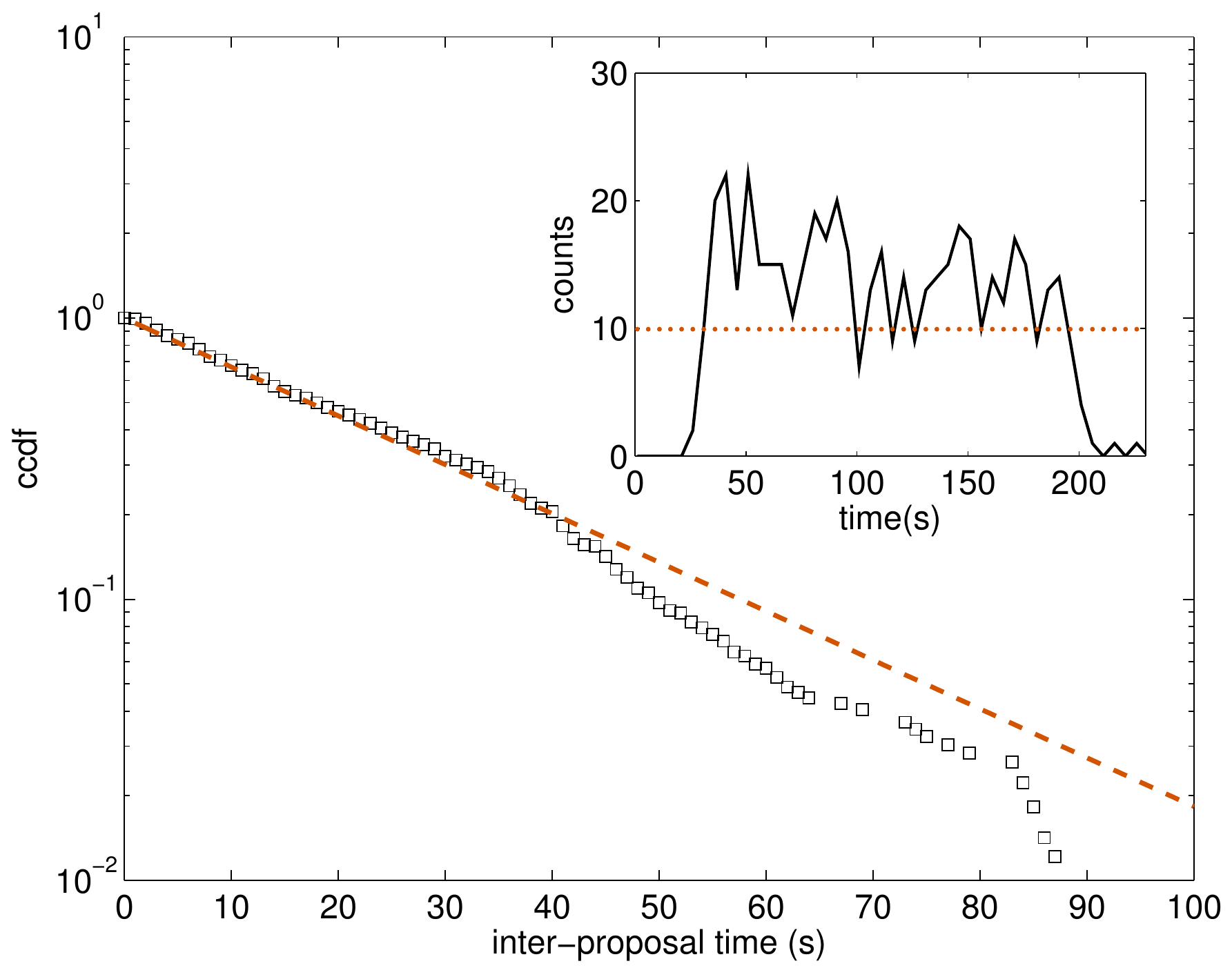}
% \vspace{5cm}
\caption{{\bf Activity of the players during the games.} 
\textbf{Main figure}: complementary cumulative distribution function (ccdf) of the inter-proposal times (bin size: $1$ s). Dashed line \revtwo{shows} an exponential fit (MLE) to the data $f(t)=\exp(-bt)$ with $b=0.036$ s$^{-1}$. 
\textbf{Inset}: Temporal distribution of the number of proposals across the games aggregated in bins of 5 s. Dotted line shows the averaged activity of 10 proposals every 5 seconds.}
\label{fig5}
\end{figure}

The activity during the game is measured by the number of complete color codes submitted by 
the players. Fig. \ref{fig5} shows how the activity of the players during the games 
fluctuates around an average activity of 10 proposals every 5 seconds. The 
complementary cumulative distribution of inter-proposal time, that is, the time 
between two consecutive proposals for the same individual, reveals an 
exponential decay of the individuals’ activity with a characteristic time of 
$28 \pm 3$ s after Maximu\revtwo{m} Likelihood Estimation (MLE) \rev{\cite{Clauset2009}.}

\begin{figure}[ht]
\includegraphics[width=\linewidth]{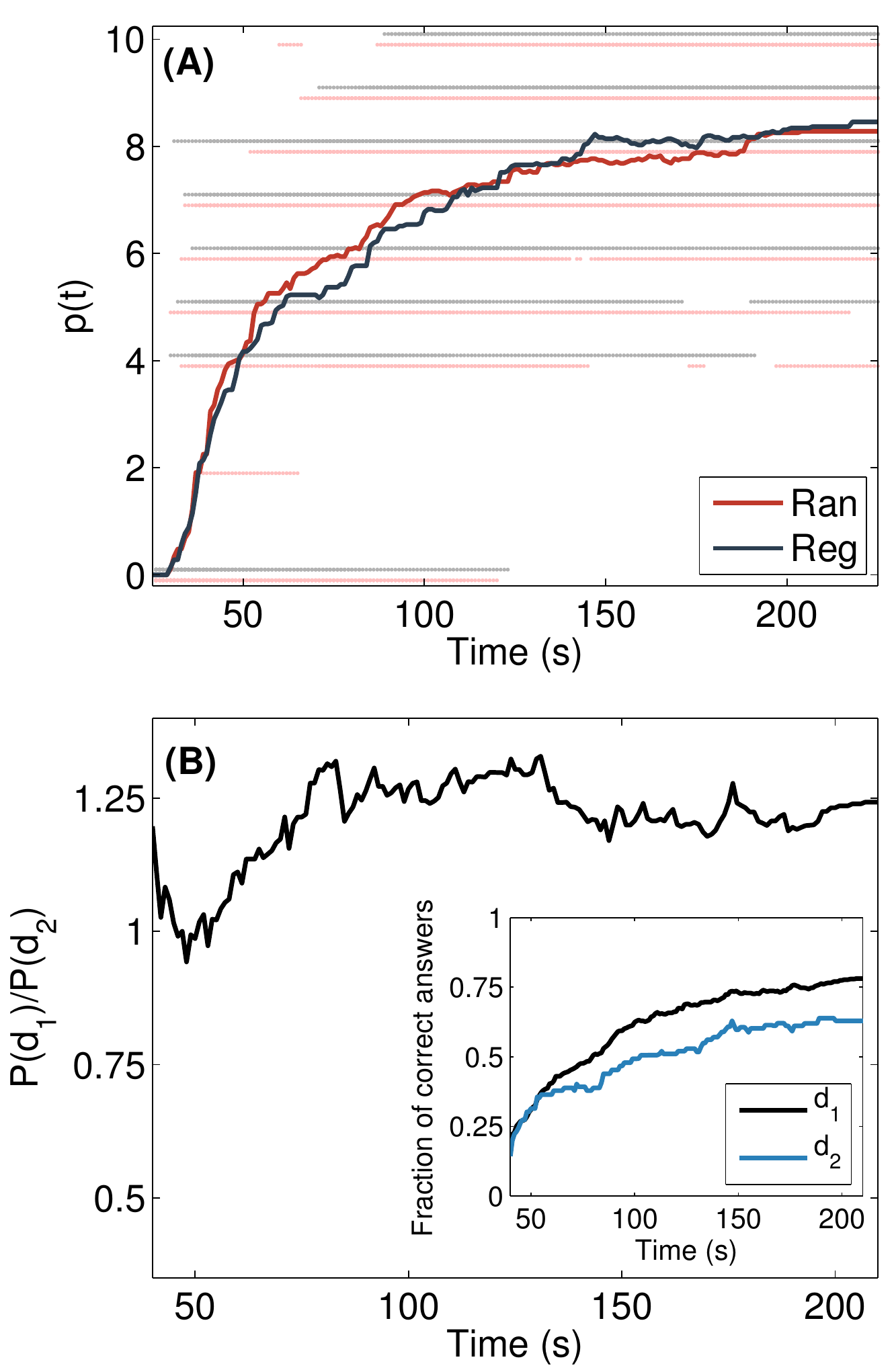}
%\vspace{3cm}
\caption{{\bf \rev{Group performance:} temporal evolution of the \rev{accuracy}.} 
\textbf{(A)} Average number of correct positions for aggregated data of the games with the same interaction topology (solid thick lines, red random networks; blue: regular lattice). 
The number of correct positions of \revtwo{the} individual players \revtwo{is} also shown (dimmed symbols). For visualization purposes a slight vertical displacement has been applied. 
\textbf{(B)} Ratio between the probabilities of correct positions at distances $d_1=1$ and $d_2=2$ from the closest source. \textbf{Inset}: Temporal evolution of the fraction of correct positions 
for distances $d_1$ and $d_2$ to the closest source. In both panels, results are aggregated over all the sessions of the experiments.}
\label{fig6}
\end{figure}

\rev{In all the games, as the time increases, the participants tend \revtwo{toward} the proposed color code. 
The hamming distance, defined as the number of differing positions between two color codes, averaged over all pairs of players at the end of each game lies in the range $(2.1,3.3)$.} %$\langle h_{ij}\rangle=0.33\pm0.01$ to $\langle h_{ij}\rangle=0.21\pm0.01$.} 
The performance of the collective is measured by the \rev{temporal} evolution of the 
\rev{accuracy} averaged over the total number of players, that is, 
$p(t)=\frac{1}{N}\sum_j\sum_i \delta_{x_j(l_i,t),c(i)}$, where \rev{$\delta_{a,b}$} is the 
Kronecker’s delta. 
Fig. \ref{fig6}A shows $p(t)$ as well as the 
individual distributions for the two networks considered in the experiment. 
The average performance of the group at the end of the \rev{games with the same interaction network is $p=8.3$.} 
\rev{The unpaired Mann-Whitney U test and the paired Wilcoxon signed rank test, computed for 
the distributions of $p(t)$ values at each time step of $1$ s indicate the absence of any 
statistical difference between the distributions of $p(t)$ in the two networks 
(average p-values $>0.4$, see Fig. S1).}

\rev{The propagation of information is investigated by measuring the relationship between the probability of 
providing a correct answer and the distance to the closest source. For a position, a node is a source if the position corresponds to 
one of the entries provided initially to the player in that node. 
The positions shown initially are correctly assigned $95\%$ of the times \revtwo{by} the players and \revtwo{practically remain unchanged} during the game, 
\revtwo{with} only 26 changes \revtwo{for these positions} out of the 1491 proposals.} 
For each player and each unknown position $l_i$, we compute the 
shortest distance to the closest source and the number of correct 
answers at a given distance $d_i$. For the two topologies and \revtwo{the} experimental conditions, any player and any 
unknown position is at maximum distance of two to the closest source. The ratio 
between the probabilities of correct answers given at distance $1$ and $2$ 
from the sources indicates that proximity to source provides an advantage 
($24.4\%$ on the average) to get the correct color (Fig. \ref{fig6}B). 

The adaptation of the players across the first and the last game in each experimental session is quantified by the averaged inter-proposal time. 
We observe similar values for the two experiments, and although the average values are similar, the last automata game shows systematically a smaller 
average inter-proposal time (Fig. S2).

\subsection*{Transition events and probabilities}
\begin{figure}[ht]
\includegraphics[width=\linewidth]{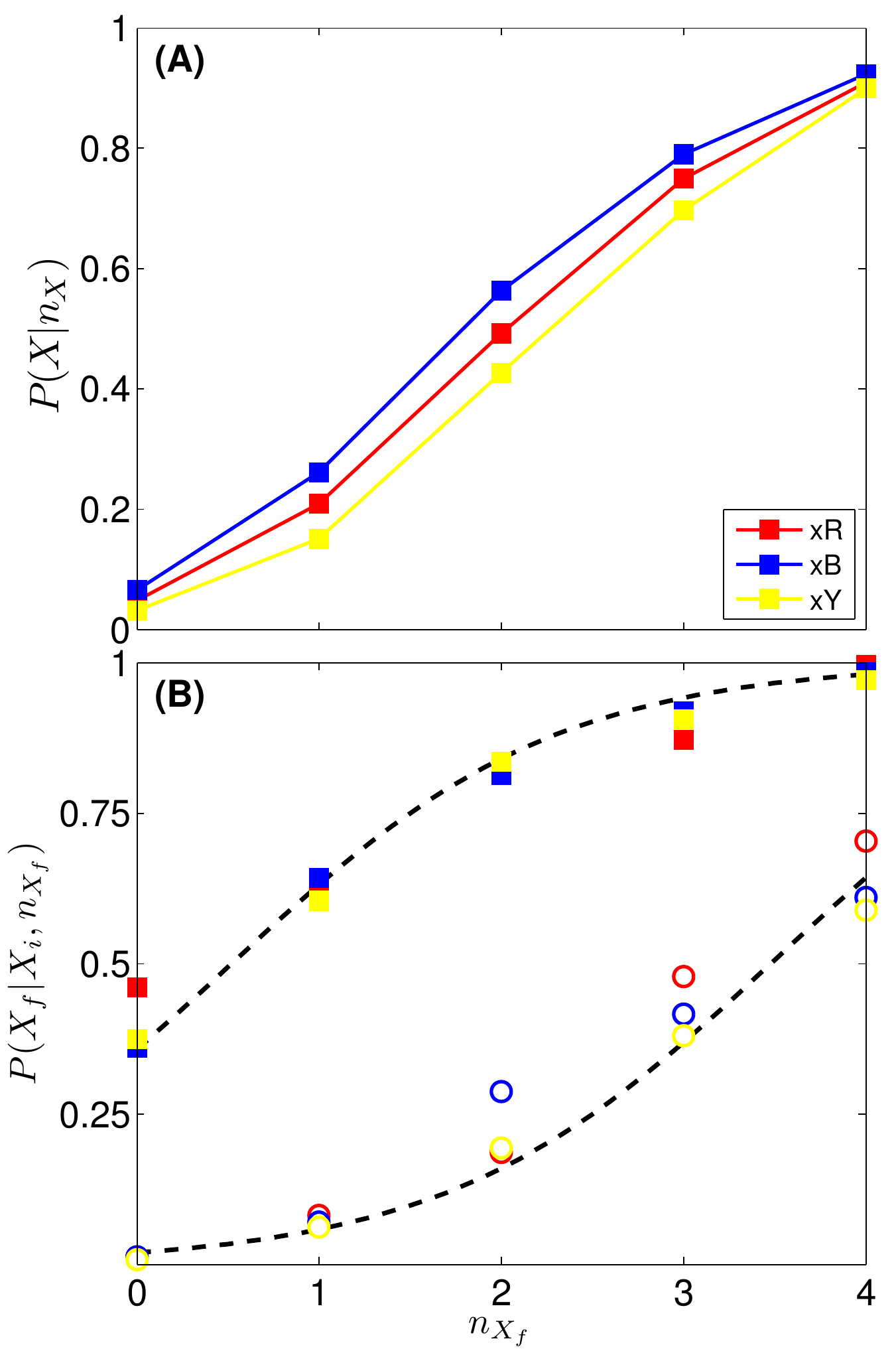}
%\vspace{3cm}
\caption{{\bf Transition Probabilities.} 
\textbf{(A)} Probability $P(X|n_{X})$ to change the unknown positions of the code from any color (R, B or Y) to a given color $X$ (R, B or Y) as a function of $n_{X}$\revtwo{, this is,} 
the number of nearest neighbors \revtwo{already} with color $X$. Results for each color are represented by the corresponding colored symbols. 
\textbf{(B)} Probabilities $P(X_f|X_i;n_{X_f})$ for the unknown positions of the code as a function of $n_{X_f}$.  
Open and solid symbols represent, for a given position, 
transitions between different colors ($\overline{X}X$) and \revtwo{between} the same color ($XX$) respectively. Results for each color are represented by the corresponding colored symbols. 
Dashed lines represent fitting Eq. (\ref{prob_h2}) to the data (see Section Bayesian update for more details).}
\label{fig7}
\end{figure}

Given the sequence of proposals of one player and the \rev{color codes} of the neighbors, we calculate 
the transition probabilities between different colors. \rev{We consider that each position is independent of each other, 
thus, we average the transition probabilities over all positions.} 
From all the possible transitions between colors, we compute the conditional 
probability to change from any color to a given color $X$, 
$P(X|n_{X})$ given $n_{X}$ neighbors in state $X$. This 
probability, shown in Fig. \ref{fig7}A, indicates that colors are practically equivalent 
for the players. This probability also points that, for a given position, as 
the number of neighbors with the same color $n_{X}$
increases, the higher the probability that the player will switch to that color 
(or remain on it).

In addition, the probability to change to a different color 
$P(X|\overline{X},n_X)$ and the probability to remain in the same color 
$P(X|X,n_X)$ measure the influence of the neighborhood to either transitions 
between colors $X\overline{X}$ or to remain in the same color (transitions 
$XX$). As Fig. \ref{fig7}B shows, both the probability to change to color $X_f$ and the 
probability to stick \rev{with} color \rev{$X_f$} increase with the number of neighbors $n_{X_f}$ 
\rev{having} color $X_f$. The neighborhood provides stronger confirmation of the 
current state of the individual than exert\revtwo{s} pressure to change it. \rev{This is, 
given a neighborhood configuration with $n_{X_f}$\revtwo{,} the probability \revtwo{for} a player to stay in her current state 
(being $X_f$) is greater than the probability of changing her state (being not $X_f$) to $X_f$.} 
Interestingly, the conviction of a player in her current state can be quantif\revtwo{ied} by the value of 
$P(X|X,0)\approx0.4$, which matches 
$P(\overline{X}|X,n^*)$ for $n^*=3$, corresponding to the neighborhood 
majority since the degree of the interaction networks was set to $k=4$.

\subsection*{Bayesian update}
We assume players update their beliefs based on their previous belief and the 
information observed from the neighborhood. Then, following Bayes theorem, 
the posterior belief of agent $i$ in state $X=R,B,Y$ conditioned to \rev{receiving a signal} $A=R,B,Y$ 
is updated as follows,

\begin{eqnarray}
P_{post}(X|A)&=&\frac{P(A|X)P_{pre}(X|A)}{\sum\limits_{\substack{x_i=R,B,Y}} P(A|x_i)P_{pre}(x_i|A)}.
\label{prob1}
\end{eqnarray}

By iterating Eq. (\ref{prob1}), the posterior belief of player $i$ after receiving $n_R$ red, $n_B$ 
blue, and $n_Y$ yellow signals is

\begin{widetext}
\begin{eqnarray}
P_{post}(X|A)&=&\frac{P(R|X)^{n_R} P(B|X)^{n_B} P(Y|X)^{n_Y} P_{pre}(X|A)}{\sum\limits_{\substack{x_i=R,B,Y}} P(R|x_i)^{n_R} P(B|x_i)^{n_B} P(Y|x_i)^{n_Y} P_{pre}(x_i|A)}.
\label{prob2}
\end{eqnarray}
\end{widetext}

Note that $\sum\limits_{\substack{A_i=R,B,Y}} P_{post}(X|A_i)=1$. 
Now, assuming that $P(A|X)=C_0$ for $A = X$  and $P(A|X)=C'$ for $A\neq X$, Eq. (\ref{prob2}) becomes

\begin{eqnarray}
P_{post}(X|A)&=&\frac{1}{1+\sum\limits_{\substack{x_i\neq X}}(\frac{C'}{C_0})^{n_X-n_{x_i}} \frac{P_{pre}(x_i|A)}{P_{pre}(X|A)}}.
\label{prob3}
\end{eqnarray}

The first hypothesis to consider is that the previous belief of the players, $P_{pre}(x_i|A)$, is not related to the actual state of the player and 
it takes the same value for the different states $X$. In this case, Eq. (\ref{prob3}) reduces to
\begin{eqnarray}
P_{post}(X|A)&=&\frac{1}{1+\sum\limits_{\substack{x_i\neq X}} s^{n_X-n_{x_i}}}.
\label{prob_h1}
\end{eqnarray}
with $s=C'/C_0$.

\begin{figure}[ht]
\includegraphics[width=\linewidth]{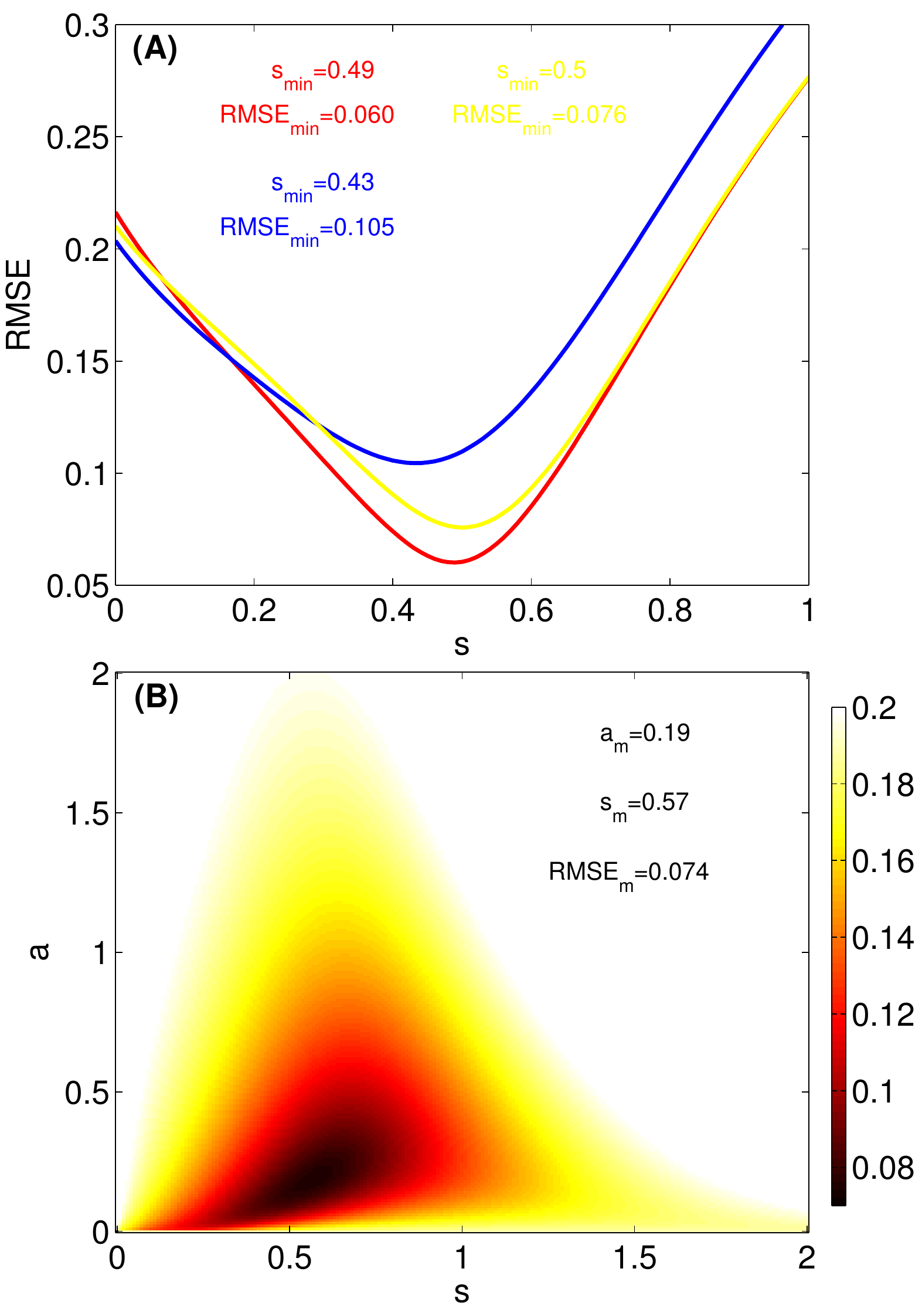}
%\vspace{3cm}
\caption{{\bf Data fitting.} 
\textbf{(A)} Root-mean-square error between the probabilities computed from experimental data (aggregated over all the experiments) and Eq. (\ref{prob_h1}). 
Results for each color are represented by the corresponding colored lines. 
\textbf{(B)} Color-coded root-mean-square error between the probabilities computed from experimental data 
(aggregated over all the colors across experiments) and Eq. (\ref{prob_h2}).}
\label{fig8}
\end{figure}

We can estimate the value of the parameter $s$ from the experimental data \revtwo{by computing} the root-mean-square error (RMSE) of the difference between 
the probabilities extracted from the experimental data and the probabilities computed using Eq. (\ref{prob_h1}). The RMSE\revtwo{, shown in} Fig. \ref{fig8}A\revtwo{, exhibits} a minimum for 
$s\approx\frac{1}{2}$ indicating that the probability \rev{that the signal is of the same value than the current state is twice the probability that the signal has a different value.}
This result is in agreement with the experimental findings \revtwo{shown in} Fig. \ref{fig7}B.

We contemplate a second hypothesis where $P_{pre}(\rev{x}_i|\rev{A})$ is not independent of the current state, so, defining $a=\frac{P_{pre}(\rev{x}_i|\rev{A})}{P_{pre}(X|\rev{A})}$ and $s=C'/C_0$, 
Eq. (\ref{prob3}) can be now written as 

\begin{eqnarray}
  P_{post}(X|A)=\begin{cases}
    \frac{1}{1+a\sum\limits_{\substack{x_i\neq X}} s^{n_X-n_{x_i}}}, & \text{if $A=X$}.\\
    \frac{1}{1+a^{-1}s^{n_X-n_{A}}+\sum\limits_{\substack{x_i\neq (X,A)}}s^{n_X-n_{x_i}}}, & \text{if $A\neq X$}.
  \end{cases}
\label{prob_h2}
\end{eqnarray}

We can now estimate the parameters $a$ and $s$ \revtwo{by} measuring the RMSE between the aggregated probabilities extracted from the experimental data and the \revtwo{probabilities computed} from Eq. (\ref{prob_h2}). 
The RMSE exhibits a minimum for the parameters $s=0.57$ and $a=0.19$ (Fig. \ref{fig8}B). The parameter $a$ \revtwo{takes into account} non-social information only. For the same number of individuals 
choosing among the options (i.e., $n_{x_1}=n_{x_2}=n_{x_3}$), $a=1$ corresponds to equal weight between the options providing a probability of $1/3$\revtwo{,} as in a random choice. 
$a<1$ implies a tendency to remain in your current state since $P(X|\rev{A=}X)$ tends to $1$ as $a$ tends to zero, \revtwo{while} for $a>1$, individuals tend to change their states more often. 
The parameter $s$ contains the social influence. For $s=1$, \revtwo{the players\textquoteright~} own opinion and the opinion of \revtwo{the} neighborhood \revtwo{are} balanced, providing a probability of $1/2$ (for $a=1$). 
However, when $s<1$ the opinion of the neighborhood counts more than \revtwo{the players\textquoteright~} own opinion\revtwo{,} while the opposite happens for $s>1$.

\subsection*{Simulations}
\begin{figure}[ht]
\includegraphics[width=\linewidth]{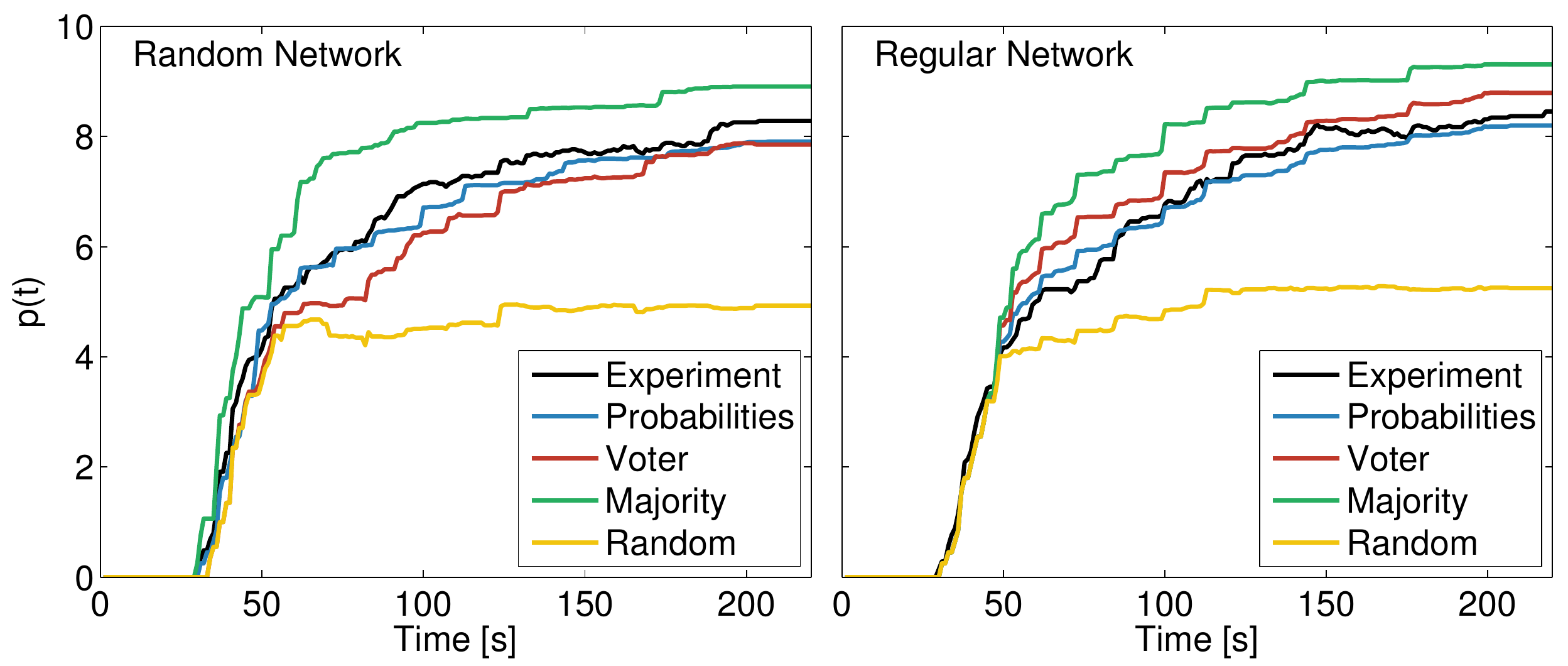}
%\vspace{3cm}
\caption{{\bf Confronting different models via simulations.} 
Temporal evolution of p(t) for different simulation models in two types of interaction networks. Probabilities corresponds to simulations using the experimentally determined probabilities. 
Experimental data is represented by the black line and it is aggregated over games with the same interaction topology. 
Simulations are averaged over 50 independent realizations.}
\label{fig9}
\end{figure}

We have performed simulations confronting various models with the experimental data\revtwo{,} where agents change their state \revtwo{according to} the experimentally determined 
transition probabilities. We have also considered other models with different updating rules. 
In the Majority rule, agents pick the 
color of the majority of their neighborhood and ties are broken by random elections among the tied options. In the Voter model, agents pick a color at random from the ones present in the 
neighborhood. In the Random model however, agents pick a color at random not necessarily present in the neighborhood. The simulations mimic the experimental conditions of the games by using the 
same number of agents, the same initial conditions, and the same temporal sequence of updates. Fig. \ref{fig9} shows the \revtwo{time} evolution of the \revtwo{performance} $p(t)$ for the different 
models and the experimental data, \rev{in both cases aggregated over games with the same interaction topology}. In \revtwo{both} networks, Majority rule outperform\revtwo{ed all} other strategies, including the one used by humans. 
In \revtwo{the} random network, Voter performs \rev{at the end of the games} as well as human strategies. In \revtwo{the} regular network, Voter performs \rev{slightly worse than} Majority\rev{, and slightly better than} human strategies. 
\revtwo{For both networks, choosing a color at random} independently of the neighbors\textquoteright ~proposals \revtwo{is} the worst strategy. Simulations also corroborate that, for the human strategy (Probabilities), there are no differences in the performance 
of the different networks considered. 

\section*{Discussion}
We developed a web-based experiment in order to assess how different interaction  topologies affect the performance of the aggregation of information and to 
\rev{investigate the \revtwo{underlying} decision making process. We found that the behavior of the individuals \revtwo{was} well captured by Bayes theory of probability update. 
Bayesian inference has been observed in human sensorimotor learning \cite{Konrad2004} and probabilistic cognition \cite{Fontanari2014} 
as well as in the behavior of many animals \cite{Valone2006,McNamara2006} (and references therein). 
The derived expression for the probabilities allows for the estimation of \revtwo{the} parameters $s$ and $a$\revtwo{,} accounting for factors such \revtwo{as}
the external social influence and the internal preference respectively. Different functional forms for these probabilities \revtwo{are reported} in the literature 
\cite{Arganda2012,Deneubourg1990,Petit2006,Eguiluz2015} 
for different animals (including humans) and behavioral situations. The parameter fitting to different experimental data show\revtwo{s} 
a high variability in the values of $s$ \rev{(social influence)} and $a$ \rev{(internal preference)}, which suggest\revtwo{s} that they are not only idiosyncratic but also task dependent. 

We found that proximity to the source provides a significant advantage in getting the correct answer. 
The propagation of information can be seen as a wave front emanating from the different sources. 
The positions shown initially are correctly assigned by the players and \revtwo{almost} do not change during the game. 
Thus, neighbors at distance $d=1$ will receive this as a constant signal compared to positions 
that are not known and consequently are changed more often. \\

With respect to the effect on the aggregation of information of the different interaction topologies considered, we found 
no significant difference on the performance of the game. This result is \revtwo{in} line with the behavior in the context of cooperation. 
In this case, players have to choose between two options: cooperate or defeat, and they get a payoff according to a payoff matrix 
and the response of their neighbors. Recent large experiments show no difference \revtwo{in the performance} at the level of number of cooperators \cite{Gracia2012}. 
However, differences in the performance depending on the network topology are reported in other works. 
For example, in search problems, efficient networks can outperform inefficient networks \cite{Mason2012} and the opposite has also been reported \cite{Lazer2007,Mason2008}. 
The reason for this contradictory result \revtwo{is} already discussed in reference \cite{Mason2012}\revtwo{,} pointing out that the level of myopia during the search plays a crucial role: 
\revtwo{in} searches performed at intermediate levels, i.e. not too close nor too far, inefficient networks can outperform efficient ones. 
It also \revtwo{established} that humans outperform simulated strategies, pointing out that\revtwo{,} for the search class of problems they have considered, agent-based simulations are not sufficiently 
sophisticated to reflect human responses \cite{Lazer2007}.} 
\revtwo{In contrast,} we found that simulations using agent-based models are \revtwo{use}ful to find a better performing strategy that solves our problem.
\revtwo{By} confronting the experimental probabilities with different numerical models based on different decision rules, we found that player\revtwo{s} did not perform optimally, \revtwo{showing} 
that the Majority rule \revtwo{outperformed players\textquoteright~ outcome} for the proposed task. 
\rev{A fundamental question that emerges from this result is why \revtwo{does} individuals use Bayesian inference despite \revtwo{the fact that}
it is not the optimal strategy\revtwo{?}. Our hypothesis is that Bayesian inference explore\revtwo{s} and maintain\revtwo{s} the diversity of options for longer time compared to 
the Majority rule that quickly narrows the set of final options. A test of this hypothesis could be done in future experiments by decreasing the initial set of positions shown \revtwo{to the players}.}\\

\rev{The adaptation is an important \revtwo{factor} to be considered in games where participants \revtwo{are made to} play several times. 
In our experimental setup we specifically designed the first and the last games to evaluate 
the adaptation of the players while minimizing the interference with the subsequent games. Players \revtwo{were} faster 
at the end of each experimental session corresponding to a decrease in the inter-proposal times.
However, in the games involving human players, we found no differences in the performance of the players. 
Thus, players familiarized with the game interface during the first game, and no further advantage \revtwo{was} obtained from acquiring more experience in the subsequent games.
\\

In all the sessions, the participants approached the correct color code. 
Whether the games lasted enough time to let the players reach consensus is an open question that needs to be addressed in future experiments.}  
\rev{Despite the results reported here, it would be interesting to consider larger networks. 
We plan to conduct more experiments with larger groups in order to address interaction networks with different neighborhood sizes 
as well as change the cognitive load of the game by varying the number of positions initially shown and the length of the code. 
It would also be interest\revtwo{ing to} study how people form consensus far from the correct solution of the problem\revtwo{,} that can be triggered by initializing incompatible targets.}

\begin{acknowledgments}
The authors acknowledge fruitful discussions with Marina Diakonova during the design of the experiment. 
The authors also acknowledge support from Spanish Ministry of Economy and Competitiveness (MINECO) and FEDER (EU) through the 
projects MODASS (FIS2011-24785) and INTENSE@COSYP (FIS2012-30634). 
T.P. acknowledges support from the program Juan de la Cierva of MINECO.
\end{acknowledgments}

\section*{Supporting Information}

% Include only the SI item label in the subsection heading. Use the 
%\nameref{label} command to cite SI items in the text.
% \subsection*{S1 Video}
% \label{S1_Video}
% {\bf Activity of the players during one game.} From top, the target color code is shown on the first row, second row is left empty (white color) for visualization purposes, 
% and in the following rows the players proposals are displayed. Initially, only the three positions seen by each player are displayed and as the players make their proposals the 
% entire color codes appear.

\subsection*{Phases of the game}
\label{S1_Text}

Each game is divided in 5 phases:
\begin{itemize}
 \item Phase 1. The goal and instructions of the game are displayed on the screen during 60 seconds.
 \item Phase 2. During 10 seconds, a different set of positions of the target color code is shown to each player. 
 The positions of the code not shown to the player are displayed as question marks over green background (see Fig. \ref{fig2_panel} A). 
 \item Phase 3. Each player has to introduce a complete color code, probably using the information provided initially, by filling the ten question marks of the empty color code.
 \item Phase 4. The player is allowed to perform as many guesses as considered necessary within the game timer countdown (225 s). 
 After introducing a first complete guess, the player is able to see 
the color codes proposed by the neighbors (see Fig. \ref{fig2_panel} C).  
 \item Phase 5. Once the timer comes to zero, the game ends and the player is rewarded with 
a score according to the number of correct colors answered.
\end{itemize}

After the individual scoring, a historical ranking of all the players with their scores is shown. 
During this stage the players are synchronized again in order to start the next game 
simultaneously.

\subsection*{Instructions to the participants}

The main goal of the game is shown at the beginning of the experiment as the following 
itemized list:

\begin{itemize}
 \item At each game a secret sequence of colors (the Secret Code) will be provided
 \item Your goal is to find all the colors of the Secret Code before the Timer (top-right) comes to zero
 \item The more colors you find, the higher will be your Score
 \item Obtain the highest Score to become the new Leader
\end{itemize}

At the beginning of each game, the following list of instructions is shown to each participant:

\begin{itemize}
 \item To begin with, we will show you part of the Secret Code
 \item We also assigned you a Team (top) that knows parts of the Secret Code
 \item Color Your Code (bottom) and Update it to see the Codes of your Team
 \item Look at the Codes of your Team to improve Your Code
\end{itemize}

A button with a summary of instructions is displayed in the main screen. That summary 
contained the following instructions:

\begin{itemize}
 \item Your goal is to find all the colors of the Secret Code before the Timer (top-right) comes to zero
 \item The more colors you find, the higher will be your Score
 \item Color Your Code (bottom) as the code initially shown
 \item You must color all the question marks of Your Code before clicking the Update button
 \item Look at the Codes of your Team (top) to improve Your Code. They contain parts of the Secret Code
\end{itemize}

\subsection*{Automata games}
In each experimental session, the first and the last game were played against automata. In order to minimize interferences with the subsequent games only two colors per position (red or blue) were used, and 
five positions of the code were initially shown to the players in these games. The first game was intended to let the players familiarize with the 
interface of the game. The last game was used to evaluate the adaptation of the players. In these games, no interconnection network was used, each player only interacted with four automata.

\subsection*{S1 Fig}
\label{S1_Fig}
\includegraphics[width=\linewidth]{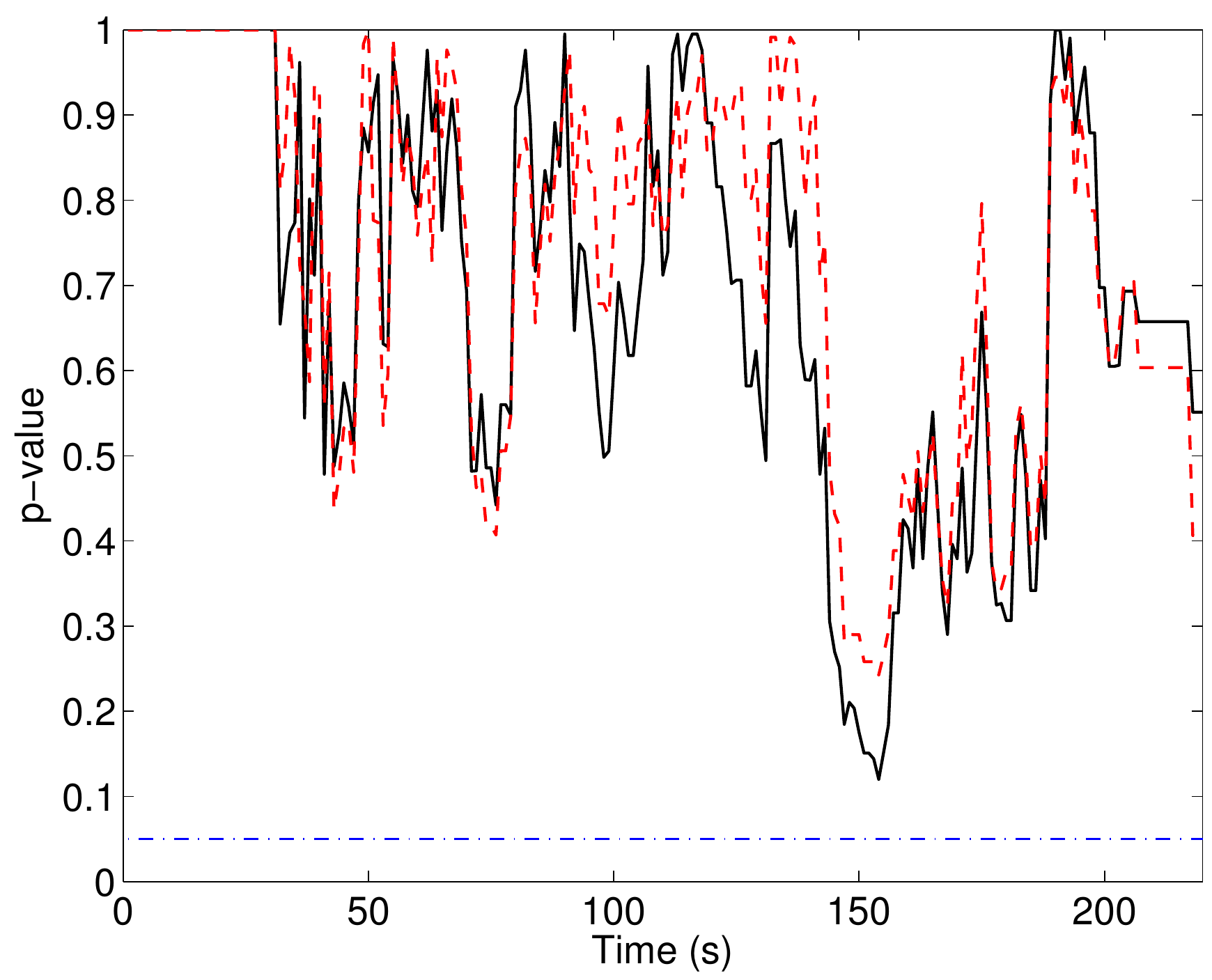}\\
{\bf Temporal evolution of the p-value of two statistical tests.}
The unpaired Mann-Whitney U test is represented by the black solid line and the paired Wilcoxon signed-rank test corresponds to the red dashed line. 
At time steps of 1 s, we applied the tests between the distributions of correct answers of the two interaction networks 
(aggregated over the different games with the same interaction topology). The dotted-dashed line indicates the $5\%$ significance level.

\subsection*{S2 Fig}
\label{S2_Fig}
\includegraphics[width=\linewidth]{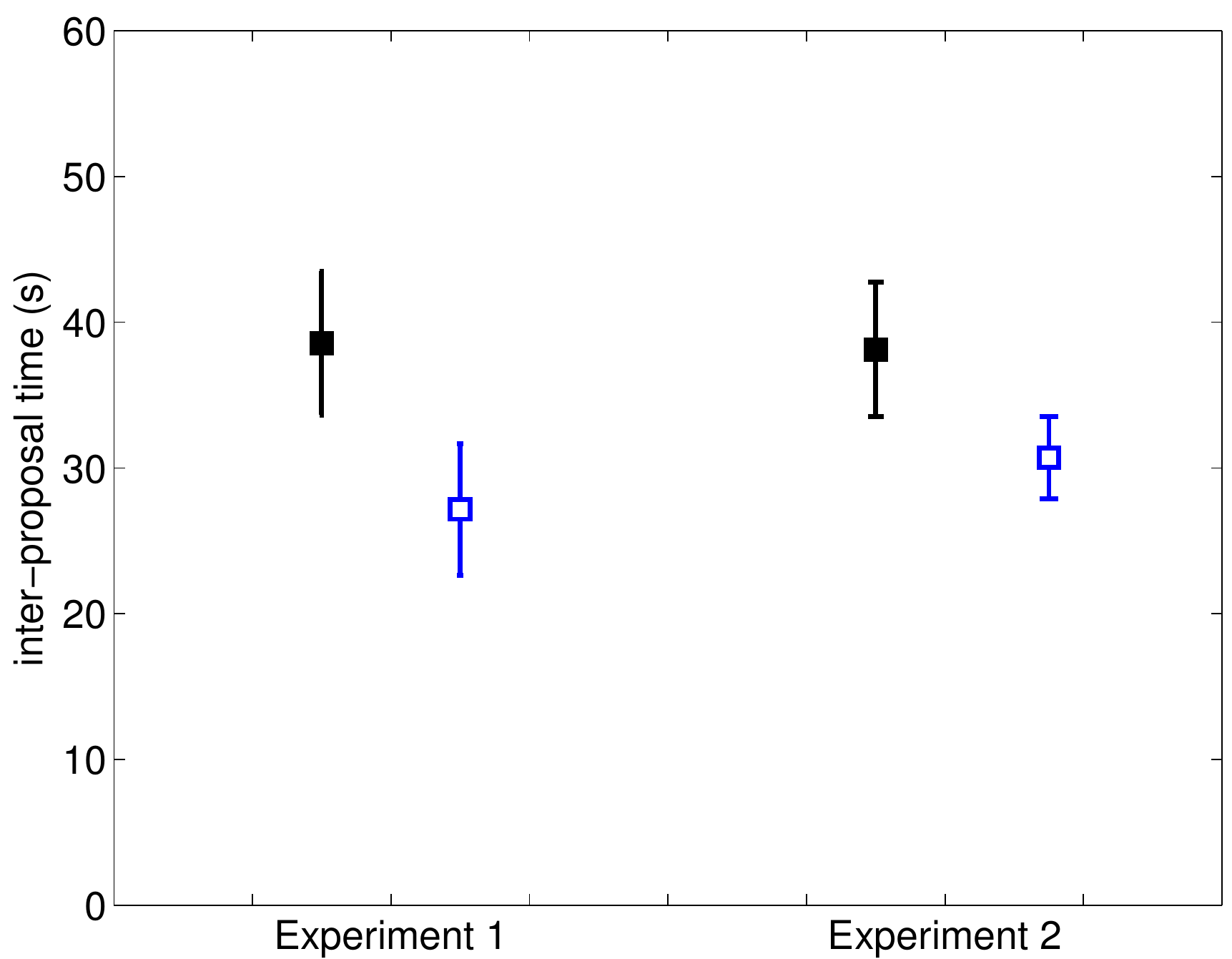}\\
{\bf Inter-proposal time for the automata games in each experiment.}
The mean inter-proposal time for the first (filled symbols) and last game (open symbols), both played against automata, for each experiment.
Bars represent the standard error.

%\subsection*{S1 Data}
%\label{S1_Data}
%{\bf Experimental data.} Data collected during experimental sessions. 


\end{document}